\documentclass[10pt,aps,prl,twocolumn,superscriptaddress,floatfix,longbibliography]{revtex4-2}

\usepackage{graphicx}
\usepackage{amsfonts}
\usepackage{amssymb}
\usepackage{amsmath}
\usepackage{txfonts}
\usepackage{xcolor}
\usepackage{cancel}
\usepackage{braket}
\usepackage{lipsum}
\usepackage[dvipsnames]{xcolor}
\usepackage{wasysym}
\usepackage[colorlinks=true, allcolors=blue]{hyperref}
\usepackage{bbold}
\usepackage{etoolbox} 
\usepackage[normalem]{ulem}
\usepackage{mathtools} 
\usepackage{multirow} 
\usepackage{hhline}
\usepackage{mathrsfs} 
\usepackage{soul}
\usepackage{array}

\usepackage{letltxmacro}
\LetLtxMacro{\oldsqrt}{\sqrt}
\renewcommand{\sqrt}[2][\mkern8mu]{\mkern-6mu\mathop{}\oldsqrt[#1]{#2}}

\begin{document}

\title{Emergent Topology from Nonlocal Electronic Correlations in One Dimension}

\author{Félix Fossati}
\email{felix.fossati@polytechnique.edu
}
\affiliation{CPHT, CNRS, {\'E}cole polytechnique, Institut Polytechnique de Paris, 91120 Palaiseau, France}

\author{Erik Linnér} 
\affiliation{International School for Advanced Studies (SISSA), via Bonomea 265, 34136 Trieste, Italy}

\author{Evgeny A. Stepanov}
\affiliation{CPHT, CNRS, {\'E}cole polytechnique, Institut Polytechnique de Paris, 91120 Palaiseau, France}
\affiliation{Coll{\`e}ge de France, Universit{\'e} PSL, 11 place Marcelin Berthelot, 75005 Paris, France}

\begin{abstract}
We demonstrate that electronic correlations in low-dimensional systems can induce topological phases starting from a topologically trivial noninteracting band structure. 
Using an advanced cluster-diagrammatic many-body approach applied to the one-dimensional extended Hubbard model, we show that tuning the nonlocal Coulomb interaction drives the emergence of bond-order-wave (BOW) and charge-density-wave (CDW) phases. 
Despite being interaction-driven and symmetry-broken, these states admit an effective low-energy single-particle description. 
In particular, the BOW phase maps onto an effective Su–Schrieffer–Heeger model, while the CDW phase, with subleading bond-order correlations, corresponds to a Rice–Mele model. 
Both phases exhibit a nontrivial topological character, manifested by the presence of localized edge states. 
Our results establish a mechanism by which nonlocal electronic correlations generate emergent topology in correlated systems.
\end{abstract}

\maketitle

The study of topological phases is one of the central themes in modern condensed matter physics. 
Their defining feature is the existence of global topological properties that are robust against local perturbations and cannot be altered without a phase transition.
A prominent example are topological insulators~\cite{PhysRevLett.95.226801, PhysRevLett.95.146802, PhysRevLett.96.106802, PhysRevLett.98.106803, PhysRevB.75.121306, PhysRevB.79.195322, bernevig_topological_2013}.
These systems possess a nontrivial topology of their noninteracting band structure, which prevents an adiabatic deformation into a topologically trivial state without closing an energy gap or breaking the relevant symmetry~\cite{Vanderbilt_2018, Cayssol_2021, 10.21468/SciPostPhysLectNotes.67}. 
Consequently, they exhibit characteristic signatures such as robust boundary states, topological invariants, and quantized response functions~\cite{RevModPhys.82.3045, annurev:/content/journals/10.1146/annurev-conmatphys-062910-140432, https://doi.org/10.1002/pssr.201206414, Ren_2016}.

The topological classification of quantum phases is traditionally formulated for noninteracting systems, where topological invariants can be constructed directly from the eigenstates of a single-particle Bloch Hamiltonian~\cite{PhysRevLett.61.2015, PhysRevLett.95.146802, doi:10.1126/science.1133734, PhysRevLett.98.106803, PhysRevB.75.121306, PhysRevB.55.1142, bradlyn2017, PhysRevX.7.041069}. 
Although this framework remains valid in the presence of weak interactions and disorder~\cite{Vanderbilt_2018}, extending it to strongly correlated systems remains a central challenge. 
This issue is particularly pronounced in one dimension (1D), where strong correlations and quantum fluctuations are greatly enhanced and give rise to collective phenomena absent in higher dimensions~\cite{giamarchi2003, PhysRevLett.95.076801, PhysRevB.74.165104}. Notable examples include Luttinger liquids~\cite{10.1143/ptp/5.4.544} and spontaneously dimerized phases~\cite{peierls1938}, including the bond-order-wave (BOW) phase~\cite{nakamura1999, PhysRevB.61.16377, PhysRevB.65.155113, PhysRevLett.88.056402, PhysRevLett.89.236401, PhysRevLett.92.236401, PhysRevB.69.035103, PhysRevLett.96.036408, PhysRevLett.99.216403, PhysRevA.90.063608, PhysRevResearch.4.L032005}, whose topological properties remain the subject of active investigation.

Extending topological classification to interacting systems requires a profound reformulation of the underlying framework. 
Several approaches have been developed to address this challenge, including formulations based on the poles and zeros of the fully interacting single-particle Green's function~\cite{PhysRevLett.105.256803, PhysRevB.83.085426, PhysRevB.84.125132, PhysRevB.96.085124, PhysRevLett.131.106601, PhysRevB.108.125115, PhysRevLett.131.236601, PhysRevB.110.L161106, PhysRevLett.133.126504, PhysRevB.86.205119, PhysRevX.2.031008}, stochastic sampling of effective single-particle Hamiltonians~\cite{Klein2021}, or local markers~\cite{PhysRevB.104.L081105}. 
Using these methods, most studies have focused on how electron-electron interactions modify or destabilize topological phases that already exist in the noninteracting limit~\cite{hohenadler2013, rachel2018, PhysRevB.82.075106, PhysRevB.85.125113, PhysRevB.85.165138, PhysRevB.87.235104, PhysRevLett.114.185701, PhysRevB.109.075148, PhysRevB.86.205119, PhysRevB.102.085122}.
In contrast, the possibility that electronic correlations alone generate a topological phase from a topologically trivial band structure has received considerably less attention~\cite{PhysRevLett.100.156401, PhysRevLett.134.053002, Herbrych2021}.
This is challenging because topological phases are typically robust against local perturbations, while addressing this problem requires incorporating strong and often nonlocal electronic interactions.

In this Letter, we demonstrate an interaction-driven mechanism through which nonlocal electronic correlations induce topological phases in the one-dimensional extended Hubbard model. 
Using an advanced cluster-diagrammatic many-body framework, we investigate symmetry-broken BOW and charge-density-wave (CDW) phases driven by nonlocal Coulomb interactions. 
We show that quantum fluctuations within these ordered states are sufficiently suppressed for the interacting self-energy to admit an effective low-energy Hamiltonian description. This enables a direct mapping of the BOW phase onto the paradigmatic Su–Schrieffer–Heeger (SSH) model~\cite{PhysRevLett.42.1698}. 
Importantly, we find that the CDW phase is not purely charge ordered but also hosts a substantial bond-order component, forming a previously unreported mixed CDW+BOW state. 
We further show that this phase maps onto an effective Rice–Mele (RM) model~\cite{Rice1982}. 
The emergent topological character of both phases is corroborated by the presence of localized edge states.

The 1D extended Hubbard model provides a minimal yet rich platform to investigate the interplay between strong correlations and topological features.
Its Hamiltonian reads:
\begin{equation*} 
H = - t \sum_{\braket{i,j}, \sigma} c_{i, \sigma}^{\dagger} c_{j,\sigma} + U \sum_{j} n_{j,\uparrow}n_{j,\downarrow} + \frac{V}{2} \sum_{\braket{i,j},\, \sigma, \sigma'} n_{i,  \sigma}n_{j, \sigma'}, 
\end{equation*}
where $c_{j, \sigma}^{(\dag)}$ is the annihilation (creation) operator for an electron at site $j$ and spin $\sigma$, and where $n_{j, \sigma} = c_{j, \sigma}^{\dagger}c^{\phantom{\dagger}}_{j, \sigma}$ is the density operator.
We consider the hopping ${t=1}$ between the nearest-neighbor lattice sites $\braket{i,j}$, and the local $U$ and nearest-neighbor $V$ Coulomb repulsion.
At half-filling, this model features a variety of quantum phases, including the Mott-insulating state driven by the local $U$~\cite{PhysRevLett.20.1445}, and CDW and BOW states driven by $V$~\cite{PhysRevB.3.2662, nakamura1999, PhysRevB.61.16377}.
The CDW phase is characterized by an alternating pattern of doubly occupied and empty sites, while the BOW state corresponds to a staggered modulation of hopping amplitudes, forming an alternating pattern of strong and weak bonds.
The position of these states on the phase diagram depends sensitively on the relative strength of $t$, $U$ and $V$~\cite{PhysRevB.61.16377, PhysRevLett.88.056402, PhysRevLett.17.1307, PhysRev.158.383, PhysRev.171.513, PhysRevB.45.4738}.

The 1D extended Hubbard model has been extensively studied at zero temperature using a combination of powerful methods, including bosonization~\cite{PhysRevLett.88.056402, PhysRevB.69.035103}, quantum Monte Carlo~\cite{PhysRevB.65.155113, PhysRevLett.92.236401}, and density-matrix renormalization group techniques~\cite{PhysRevB.61.16377, PhysRevLett.99.216403}. 
In the zero-temperature phase diagram, the BOW phase emerges at weak to intermediate coupling, enclosed by CDW and antiferromagnetic (AFM) phases. 
At large interaction strengths, a direct AFM–CDW transition line appears, as the BOW phase is suppressed. 
Despite extensive studies at zero temperature, the interplay between local correlations and the CDW and BOW phases driven by nonlocal interactions at finite temperatures remains largely unexplored, primarily due to the difficulty of incorporating nonlocal Coulomb interactions in advanced finite-temperature numerical methods. Moreover, the BOW phase cannot be captured within single-site approaches, such as (extended) dynamical mean-field theory (EDMFT)~\cite{RevModPhys.68.13, PhysRevB.52.10295, PhysRevLett.77.3391, PhysRevB.61.5184, PhysRevLett.84.3678, PhysRevB.63.115110}, as its order parameter is intrinsically nonlocal and requires at least a dimer-based cluster treatment.

To address this challenge, we employ a recently developed cluster-diagrammatic approach~\cite{dhm8-5ss6} based on the dual triply irreducible local expansion (\mbox{D-TRILEX}) framework~\cite{PhysRevB.100.205115, PhysRevB.103.245123, 10.21468/SciPostPhys.13.2.036}. 
This method formulates a diagrammatic expansion around an arbitrary interacting reference system~\cite{DF_Brener, StepanovHDR}, thereby incorporating correlation effects beyond the reference problem. 
Here, the reference system is chosen as a two-site DMFT impurity cluster (dimer), which enables a nonperturbative treatment of local and nearest-neighbor correlations essential for capturing bond-order fluctuations. 
To ensure a consistent treatment of the nonlocal Coulomb interaction across different regions of the phase diagram, the dimer problem is solved at ${V=0}$, with the nonlocal interaction subsequently incorporated through the diagrammatic part of the \mbox{D-TRILEX} formalism~\cite{10.21468/SciPostPhys.13.2.036, StepanovHDR}. 
The ability to perform a diagrammatic expansion around an interacting dimer reference system is a key advantage of the present approach, enabling calculations directly within the spontaneously symmetry-broken BOW and CDW phases. 
To this end, we extend the implementation of Ref.~\cite{dhm8-5ss6} to allow for the self-consistent inclusion of symmetry-breaking fields. Details of the method are provided in the Supplemental Material (SM)~\cite{SM}.
The cluster \mbox{D-TRILEX} results are further compared to single-site calculations, highlighting the importance of non-perturbative short-range correlations in the dimer formulation.

\begin{figure}[t]
    \centering
    \includegraphics[width=\linewidth]{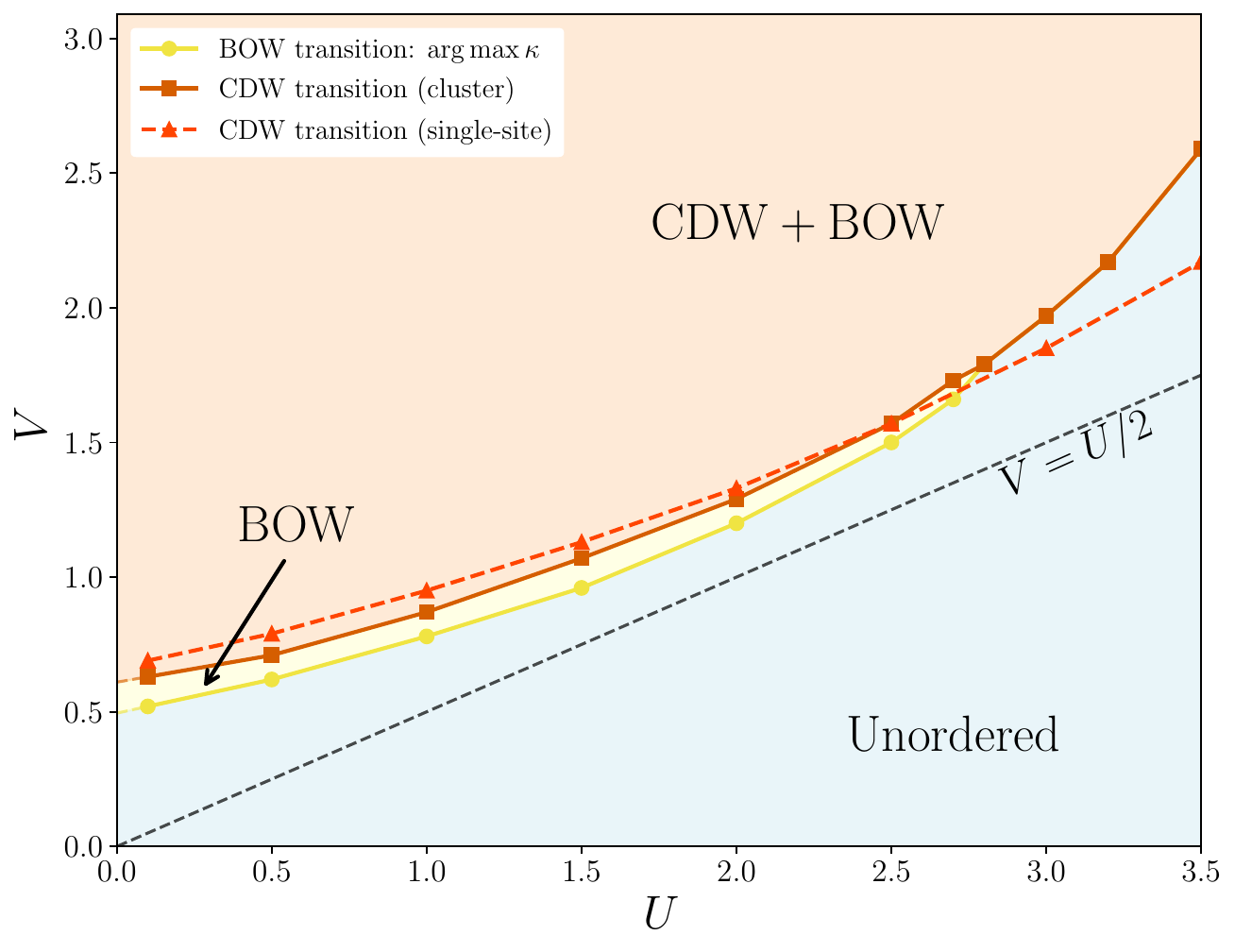}
    \caption{ The $U$-$V$ phase diagram of the half-filled 1D extended Hubbard model obtained at $T=0.1$ using \mbox{D-TRILEX}.
    The CDW phase boundaries calculated within the single-site and dimer frameworks are shown in dashed and solid red curves, respectively.
    The mean-field estimate for the CDW transition ${V=U/2}$ is given by the dashed black line.
    The BOW phase boundary obtained in the dimer calculation is depicted by the yellow curve.
    \label{fig:phase-diagram}}
\end{figure}

The resulting phase diagram of the 1D extended Hubbard model in the $(U,V)$ plane obtained at finite temperature ${T=0.1}$ is shown in Figure~\ref{fig:phase-diagram}. 
It displays three phases: a disordered phase (blue), a BOW phase (yellow), and a CDW phase (orange). 
The critical interaction strengths for the onset of the BOW ($V_c^{\rm BOW}$, yellow curve) and CDW ($V_c^{\rm CDW}$, red curves) states increase monotonically with $U$.
This behavior reflects the suppression of charge fluctuations by increasing on-site repulsion, which in turn requires stronger nearest-neighbor interactions to stabilize ordered states. 
We find that the CDW phase boundary lies systematically above the mean-field estimate ${V=U/2}$~\cite{OLES1984323, PhysRevB.39.9397, PhysRevB.42.465} (black dashed line), as expected since mean-field theory neglects fluctuations that generally reduce the stability of ordered phases.

The critical interaction strengths $V_c^{\mathrm{BOW}}$ and $V_c^{\mathrm{CDW}}$ are determined differently in the single-site and dimer formulations due to the spatial constraints imposed by their respective reference systems. 
The single-site \mbox{D-TRILEX} approach preserves the symmetries of the lattice Hamiltonian and therefore cannot access symmetry-broken phases directly. 
Instead, phase transitions are identified through instabilities of the normal state signaled by divergences of the corresponding susceptibilities. 
For the CDW transition, this occurs in the static charge susceptibility ${X^{\rm ch}(q,\omega=0)=\langle n_{q,0}\,n_{-q,0}\rangle-\langle n\rangle^2}$ at the ordering wave vector ${q=\pi}$. 
The resulting CDW phase boundary obtained from the single-site calculation is shown in Figure~\ref{fig:phase-diagram} as the dashed red curve. 
In contrast, identifying the BOW transition within the single-site framework is considerably more challenging. 
Because the BOW order parameter is intrinsically nonlocal, the corresponding susceptibility has a generalized two-particle structure involving nonlocal vertex corrections, whose consistent treatment remains a major challenge for contemporary diagrammatic approaches.

\begin{figure}[t]
    \centering
    \includegraphics[width=\linewidth]{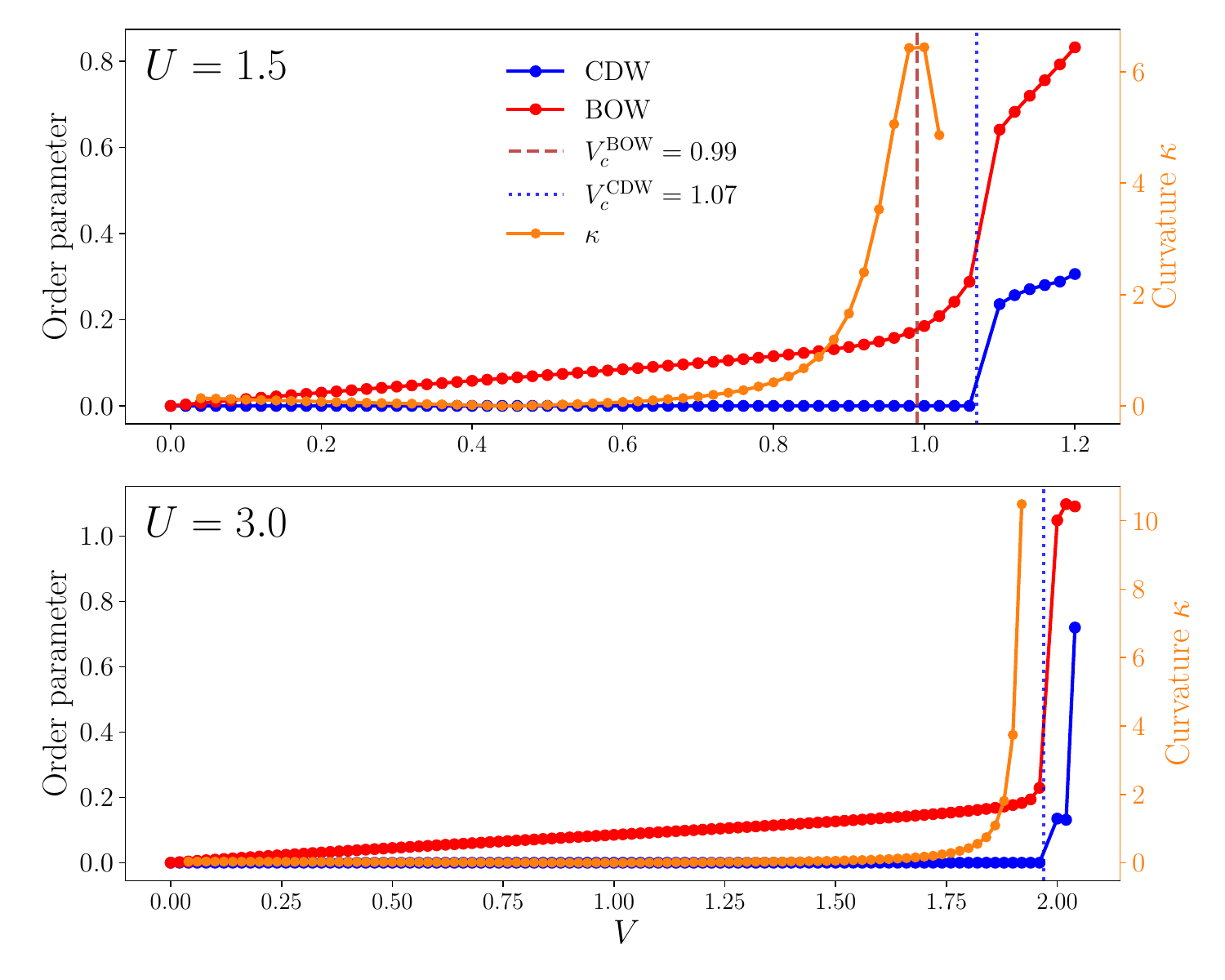}
    \caption{Evolution of the CDW (blue markers) and BOW (red markers) order parameters with nonlocal interaction $V$.
    The results are obtained within the dimer \mbox{D-TRILEX} approach for ${U=1.5}$ (top panel) and ${U=3.0}$ (bottom panel).
    The curvature $\kappa$ of the BOW curve is shown in orange.
    $V_{c}^{\text{CDW}}$ (dotted vertical blue line) marks the CDW onset.
    $V_{c}^{\text{BOW}}$ (dashed vertical red line) corresponds to the maximum of $\kappa(V)$ before the CDW sets in and is present only at ${U=1.5}$.    
    }
    \label{fig:cdw-bow-v}
\end{figure}

Within the dimer formulation, the CDW and BOW phases become directly accessible through the self-consistent inclusion of the corresponding symmetry-breaking fields~\cite{SM}. 
The CDW order parameter is defined as the staggered charge density:
$O_{\mathrm{CDW}} = \frac{1}{N}\sum_{j,\sigma}(-1)^{j}\langle n_{j,\sigma} \rangle$, 
where $N$ is the number of lattice sites.
The BOW phase is characterized by bond dimerization. 
Since the bare hopping amplitude is translationally invariant, the BOW order parameter is obtained from the renormalization of the hopping induced by electronic correlations. 
Within the dimer framework, this renormalization corresponds to the difference between the intra- and inter-dimer nearest-neighbor self-energies at the lowest Matsubara frequency $\nu_0$:
$O_{\mathrm{BOW}} = \frac12\text{Re}\,[\Sigma_{\mathrm{intra}}(\nu_0) - \Sigma_{\mathrm{inter}}(\nu_0)]$.
The determination of the BOW phase boundary requires additional care. 
Because the dimer reference system treats intra- and inter-cluster correlations differently, the former non-perturbatively within DMFT and the latter diagrammatically, the calculation is effectively performed in the presence of an intrinsic BOW symmetry-breaking field. 
As a result, $O_{\mathrm{BOW}}$ remains finite even outside the ordered phase. 
To identify the genuine interaction-driven transition, we adopt an approach analogous to locating magnetic phase transitions in the presence of an external field and define $V_c^{\mathrm{BOW}}$ from the maximum of the order-parameter curvature $\kappa(V)$ (see Figure~\ref{fig:cdw-bow-v}). This criterion identifies the onset of BOW beyond the quasi-linear background induced by the dimer reference system.

Figure~\ref{fig:cdw-bow-v} shows the behavior of $O_{\mathrm{CDW}}$ and $O_{\mathrm{BOW}}$ as a function of $V$ for two values of the local interaction, $U=1.5$ and $U=3.0$. 
In both cases, the CDW order parameter develops at the critical value $V_c^{\rm CDW}$ upon increasing $V$, while $O_{\mathrm{BOW}}$ exhibits an approximately linear behavior for $V<V_c^{\rm BOW}$ discussed above. 
For ${U=1.5}$, the BOW transition occurs prior to the onset of CDW order, whereas for ${U=3.0}$ a pure BOW phase is not stabilized.
Notably, for both values of $U$ the onset of CDW order is accompanied by a pronounced enhancement of the BOW order parameter, which shows a discontinuous increase at $V_c^{\rm CDW}$. 
This indicates that, in the CDW phase, bond order is not suppressed but instead coexists with charge order. 
Physically, while the CDW phase breaks bond-centered inversion symmetry, the simultaneous enhancement of $O_{\mathrm{BOW}}$ reflects an additional breaking of site-centered inversion symmetry. 
This coexistence of CDW and BOW orders constitutes a key ingredient in the formation of nontrivial topological features, as discussed below.
Accordingly, the orange region in the phase diagram (Figure~\ref{fig:phase-diagram}) is identified as a mixed CDW+BOW phase.

Comparison between the single-site and dimer \mbox{D-TRILEX} results for the CDW transition reveals two distinct regimes. 
At weak to intermediate coupling (${U\lesssim2.5}$), the critical lines obtained from the two approaches lie in close proximity (dashed and solid red lines in Figure~\ref{fig:phase-diagram}). 
This indicates that nonlocal correlations remain perturbative and are adequately accounted for by diagrammatic corrections within \mbox{D-TRILEX}. 
The single-site transition line lies slightly above the cluster result, with a small offset attributable to preformed bond dimerization induced by the dimer reference system, which mildly favors charge ordering.
For larger interactions (${U\gtrsim3.0}$), the two approaches deviate significantly: the single-site CDW boundary continues its quasi-linear increase with increasing $U$, whereas the cluster result bends upward, yielding substantially larger values of $V_c^{\mathrm{CDW}}$. 
This signals a regime where the non-perturbative nonlocal correlations, captured by the dimer reference system, become essential for an accurate description of the phase boundary.

This difference can be traced to the emergence of strong short-range spin correlations at larger $U$. 
As the on-site repulsion increases, local magnetic moments form~\cite{PhysRevB.105.155151, PhysRevLett.132.236504}, and neighboring spins develop pronounced singlet correlations. 
These correlations compete with charge order, since the singlet formation is incompatible with the charge disproportionation characteristic of the CDW state. 
The dimer reference system captures this competition explicitly through its exact treatment of intra-cluster correlations, whereas the single-site approach incorporates them only perturbatively via diagrammatic corrections. 

This interpretation is supported by the behavior of the nearest-neighbor static spin susceptibility, ${X^{\rm sp}_{\langle{}ij\rangle} = \langle m^{z}_{i} \, m^{z}_{j} \rangle}$, where ${m^{z} = n_{\uparrow} - n_{\downarrow}}$ is the local magnetization density.
The magnitude of the intra-cluster susceptibility $X^{\rm sp}_{\rm intra}$ increases with increasing $U$, whereas the inter-cluster one $X^{\rm sp}_{\rm inter}$ remains nearly unchanged.
At $V=0$, the two are comparable for $U=1.5$, with $X^{\rm sp}_{\rm intra}\simeq X^{\rm sp}_{\rm inter}\simeq 0.6$, while at $U=3.0$ they differ substantially, reaching $X^{\rm sp}_{\rm intra}\simeq1.6$ and $X^{\rm sp}_{\rm inter}\simeq0.5$. This demonstrates that increasing the on-site repulsion enhances the non-perturbative short-range singlet correlations captured by the dimer reference system.
The same conclusion is supported by the dependence on the nonlocal interaction $V$. 
For all values of $U$, $X^{\rm sp}_{\rm intra}$ remains nearly constant as a function of $V$ until the system approaches the CDW transition, whereas $X^{\rm sp}_{\rm inter}$ is reduced by approximately a factor of two between $V=0$ and the onset of the CDW state. 
This behavior indicates that the short-range singlet correlations encoded in the dimer reference system are remarkably robust, while the inter-cluster spin correlations are substantially more susceptible to the perturbative treatment of the nonlocal interaction.

The same mechanism governs the evolution of the BOW phase, which at intermediate couplings forms a distinct region preceding the CDW state, but merges with the CDW transition for larger values of ${U\gtrsim2.8}$ (yellow curve in Figure~\ref{fig:phase-diagram}). 
This behavior reflects the competition between short-range correlations: while the BOW phase arises from a spontaneous dimerization of the bonds, strong local singlet formation tends to homogenize bond correlations. 
As a result, increasing $U$ enhances local singlet correlations and progressively shifts the BOW instability to larger values of $V$, eventually pushing it beyond the CDW transition, so that bond dimerization develops only within the CDW phase.

The detected BOW phase, characterized by an alternating pattern of strong and weak bonds, is naturally reminiscent of the SSH model. 
In contrast, the discovered CDW+BOW state, which exhibits a subleading bond-order component in addition to a staggered charge modulation, closely resembles the RM model. 
This analogy motivates a systematic analysis of the topological properties of these interaction-driven phases. 
To this end, we employ the topological Hamiltonian approach~\cite{PhysRevX.2.031008, wang2013, PhysRevB.104.195125, PhysRevB.104.085116, PhysRevB.107.245145}, which maps the interacting system onto an effective noninteracting Hamiltonian defined as $H_{\mathrm{eff}}(k) = \varepsilon_{k}+\text{Re}\,\Sigma(k,\nu=0)$, where $\varepsilon_k$ is the noninteracting dispersion and $\Sigma$ is the many-body self-energy.
This mapping is justified in the present case by the strong suppression of dynamical correlations in both ordered phases. 
In particular, the imaginary part of the self-energy vanishes at low frequencies, ${\text{Im}\,\Sigma(k,\nu\to0)\to0}$, allowing for a quasistatic description (see SM~\cite{SM}).
As a result, the topological characterization reduces to that of an effective band structure, to which standard noninteracting invariants can be applied.

The effective Hamiltonian extracted from the dimer \mbox{D-TRILEX} calculations can be written in a Bloch form using the Pauli matrices ${\{\sigma^x, \sigma^y, \sigma^z\} }$:
\begin{equation*}
    H_{\mathrm{eff}}(k) = \begin{pmatrix}
        w_k & u_k + v_k e^{-ik} \\
        u_k + v_k e^{ik} & -w_k
    \end{pmatrix},
\end{equation*}
where $u_k$, $v_k$, and $w_k$ are weakly momentum-dependent effective parameters. 
In the BOW phase, the physics is governed by the bond parameters $u_k$ and $v_k$, whose difference encodes the dimerization, while ${w_k=0}$, reflecting the absence of charge modulation. 
The resulting effective model is therefore of SSH type, albeit with weak momentum-dependent corrections. 
In contrast, in the CDW+BOW phase, a finite $w_k$ signals the presence of charge order in addition to bond modulation. 
The effective description then corresponds to a RM model with weak momentum-dependent parameters, consistent with the RM model being a staggered-charge extension of the SSH chain~\cite{Rice1982}.

\begin{figure}[t]
    \centering    \includegraphics[width=\linewidth]{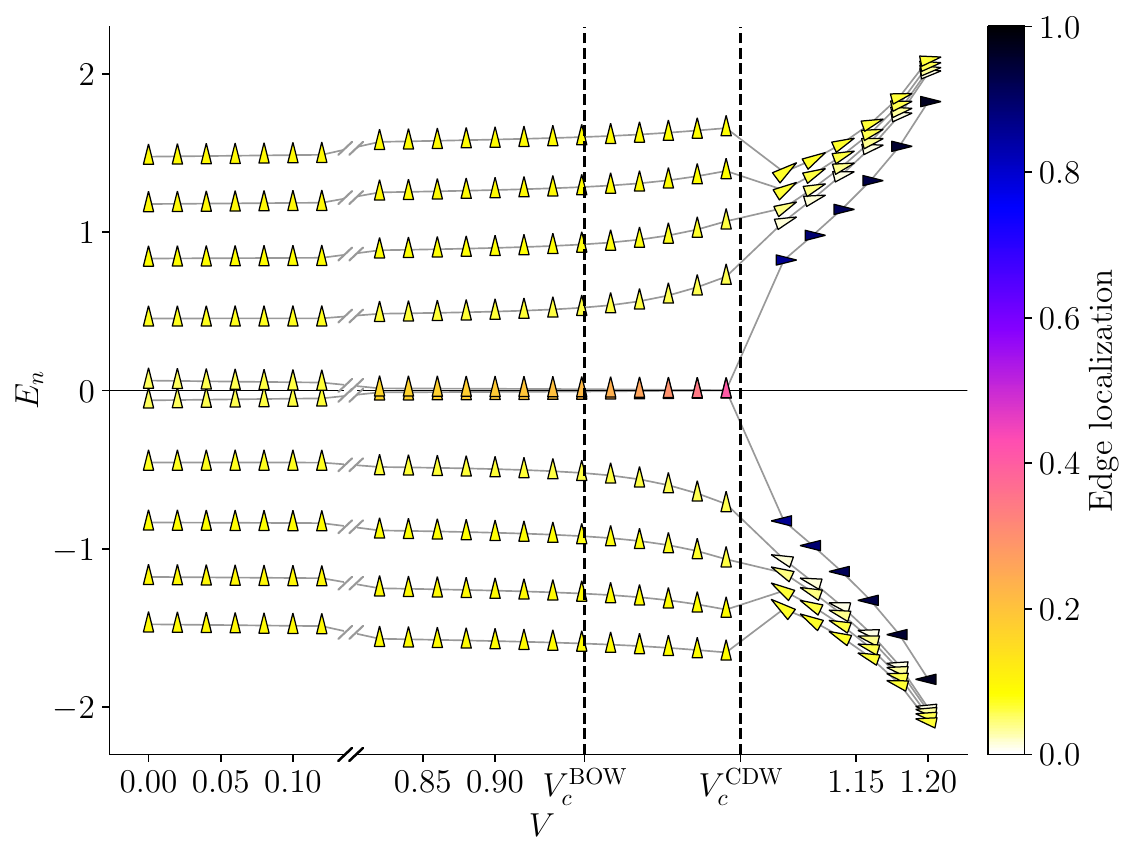}
    \caption{
    Energy spectrum of $H_{\mathrm{eff}}$ for a 64-site chain with open boundary conditions obtained as a function of $V$ at ${U=1.5}$. 
    For clarity reasons, only a representative subset of the low-energy levels are shown and intermediate values of $V$ are removed.
    Each triangle marker represents a single eigenstate: its orientation
    encodes the sublattice polarization $\mathcal{P}^{(n)}$, and its color encodes the edge-localization weight $|\phi^{(n)}|^2_{\mathrm{edge}}$, from bulk-extended (yellow) to boundary-localized (red to black).
    Vertical dashed lines mark $V_c^{\mathrm{BOW}}$ and $V_c^{\mathrm{CDW}}$.
    \label{fig:ribbon-spectrum}}
\end{figure}

The topological character of the BOW and CDW+BOW phases can be probed through the eigenstates of the effective Hamiltonian on a finite open chain. 
To this end, we perform an exact diagonalization of $H_{\mathrm{eff}}$ at ${U=1.5}$ for a chain of ${L=64}$ sites with open boundary conditions, parametrized with unpaired site at the edges. 
To characterize the eigenstates $|\phi^{(n)}\rangle$, we introduce the edge-localization weight ${|\phi^{(n)}|^2_{\mathrm{edge}} = \sum_{j \in \mathrm{edges}} |\phi^{(n)}_j|^2}$, measuring the spectral weight on the two outermost sites $j$, and the sublattice polarization ${\mathcal{P}^{(n)} = \sum_j (-1)^j |\phi^{(n)}_j|^2}$, which distinguishes states predominantly residing on one sublattice from the other. 
Both quantities are encoded in Figure~\ref{fig:ribbon-spectrum}, which shows the evolution of the eigenvalue spectrum with increasing $V$.

For $V<V_c^{\mathrm{BOW}}$, the spectrum contains no in-gap states and all eigenstates remain bulk-like (yellow markers) with negligible sublattice polarization, $|\mathcal{P}^{(n)}|\approx0$ (triangles pointing upward). 
As $V$ approaches $V_c^{\mathrm{BOW}}$, two states detach from the bulk continuum and move toward zero energy. 
Upon entering the BOW phase, these states become strongly localized at the chain boundaries (red markers) while remaining essentially unpolarized, closely resembling the topological zero-energy edge modes of the SSH model.
A qualitatively different behavior emerges at the CDW transition. 
For ${V>V_c^{\mathrm{CDW}}}$, the spectrum reorganizes into two bands separated by a polarization gap, with states predominantly residing on opposite sublattices (triangles pointing left/right). 
The boundary-localized modes persist but are shifted away from zero energy and acquire a finite sublattice polarization (black markers). This behavior is characteristic of the RM model, where inversion-symmetry breaking removes the topological protection of the SSH edge states. 
Nevertheless, the strongly localized states closest to the gap edge provide a clear spectral signature of the enhanced bond dimerization and the mixed CDW+BOW character of the ordered phase.

To conclude, we have demonstrated that nonlocal electronic correlations can drive the formation of topological phases from a topologically trivial band structure in the 1D extended Hubbard model. 
Using a cluster-diagrammatic extension of the \mbox{D-TRILEX} framework, that provides a consistent treatment of the ordered phases, we established the finite-temperature phase diagram and identified both BOW and CDW instabilities. 
Remarkably, we find that the CDW phase hosts a substantial bond-order component, forming a mixed CDW+BOW state that, to the best of our knowledge, has not been reported previously.
The suppression of dynamical correlations within these ordered phases enables an effective low-energy Hamiltonian description based on the interacting self-energy. Within this framework, the BOW and CDW+BOW phases emerge as interaction-induced realizations of the SSH and RM models, respectively. Their nontrivial character is manifested by the appearance of strongly localized edge modes in the effective spectrum.
More broadly, our work demonstrates that nonlocal electronic correlations can serve not only to modify existing topological phases, but also to generate them from an initially trivial electronic structure, opening a route toward interaction-driven topological quantum matter beyond the conventional band-theory paradigm.

\begin{acknowledgments}
F.F. and E.A.S. acknowledge support from TGCC-GENCI through the A0180901393 project.
E.L. acknowledge financial support from MUR through the PRIN 2020 (Prot. G93C22000430006), from National Recovery and Resilience Plan (PNRR) MUR (Prot. PE0000023-NQSTI), financed through Next Generation EU by the European Union, and from FVG through the Complementary Operational Program (POC) 2014-2020 (Prot. G93C25000880002).
E.A.S. also acknowledges support from ANR JCJC grant ``ELECTRO'', ANR-25-CE30-7064.
\end{acknowledgments}

\bibliography{references}

\end{document}


\title{Supplementary Material\\[3pt]
Emergent Topology from Nonlocal Electronic Correlations in One Dimension}

\author{Félix Fossati}
\email{felix.fossati@polytechnique.edu}
\affiliation{CPHT, CNRS, {\'E}cole polytechnique, Institut Polytechnique de Paris, 91120 Palaiseau, France}

\author{Erik Linnér}
\affiliation{International School for Advanced Studies (SISSA), via Bonomea 265, 34136 Trieste, Italy}

\author{Evgeny A. Stepanov}
\affiliation{CPHT, CNRS, {\'E}cole polytechnique, Institut Polytechnique de Paris, 91120 Palaiseau, France}
\affiliation{Coll{\`e}ge de France, Universit{\'e} PSL, 11 place Marcelin Berthelot, 75005 Paris, France}

\maketitle

\section{Local-field surrogate for the cluster self-consistency}
\label{sec:sm_local_field}

\subsection{Methodology}

The cluster extension of the diagrammatic \mbox{D-TRILEX} approach~\cite{dhm8-5ss6} is based on a dimer DMFT reference problem. 
This reference system treats local and nearest-neighbor correlations exactly, while correlations beyond the cluster are incorporated in a mean-field fashion via a frequency-dependent hybridization function $\Delta_\nu$, determined self-consistently through the DMFT condition. 
The remaining correlation effects are then included perturbatively within the \mbox{D-TRILEX} diagrammatic expansion.
Once the diagrammatic calculation is converged, the lattice Green's function, summed over momenta,
$G_{\nu} = \frac{1}{N}\sum_{\mathbf k} G_{\mathbf k \nu}$,
generally differs from the reference Green’s function $g_\nu$, i.e., $g_\nu \neq G_{\nu}$. 
Since the purpose of the reference system is to capture correlations of the lattice model within the cluster as accurately as possible, a significant mismatch between $g_\nu$ and $G_{\nu}$ would indicate that the chosen reference problem is not optimal for the lattice system under consideration.

The usual way to improve the reference problem is to impose an \emph{outer} self-consistency~\cite{PhysRevB.77.033101, PhysRevB.93.045107} that updates the hybridization function $\Delta_\nu$ after each diagrammatic iteration until $g_\nu \overset{!}{=} G_\nu$ is satisfied. 
We note that the hybridization function fulfilling this outer self-consistency condition no longer satisfies the DMFT self-consistency condition. 
Such a calculation is prohibitively expensive numerically in a multiband case, since each update of the hybridization function requires recomputing both single-particle quantities and, more importantly, two-particle quantities such as vertex functions and susceptibilities needed for the diagrammatic expansion.

In the case of the BOW and CDW phases considered here, the outer self-consistency can be imposed approximately, as these phases are characterized by a static symmetry-breaking field $h$. 
This field is represented by a Hermitian matrix in cluster space and acts in the charge (spin-symmetric) channel. 
Its components have a direct physical interpretation: the uniform diagonal component corresponds to a chemical-potential shift ($\delta\mu = -h$), the staggered diagonal component acts as the on-site potential conjugate to CDW order, and the off-diagonal inter-site component corresponds to the bond field conjugate to BOW order. 
Thus, a single static matrix $h$ encodes the full set of order parameters of interest.
The field $h$ enters the formalism as a static contribution to the hybridization function, which is updated under outer self-consistency according to $\Delta_\nu \rightarrow \Delta_\nu + h$.

The value of field $h$ is usually not very large, but introducing this field in the reference problem triggers spontaneous symmetry breaking, resulting in a substantial contribution to the self-energy.
Therefore, one can assume that the hybridization function changes only slightly within the outer self-consistency and this change affects mainly the single-particle quantities. 
This allows us to do calculations without updating the reference problem and to calculate the self-energy correction using the linear response approximation. 
The linear response of the reference self-energy to the applied field can be obtained as follows:
\begin{equation}
\partial\Sigma_{\nu,\,l_1 l_2}
= \partial\left(g^{-1}_{0,\nu} - g^{-1}_{\nu}\right)
= -h_{l_1l_2} - \partial\left(g^{-1}_{\nu}\right)
= \sum_{l_3,l_4}\left(\lambda^{\mathrm{ch}}_{\nu,\omega=0,\,l_1 l_2,\,l_3 l_4} - \delta_{l_1,l_3}\delta_{l_2,l_4}\right) h_{l_3 l_4} 
= \sum_{l_3,l_4} \bar\lambda^{\mathrm{ch}}_{{\nu,\omega=0},\,l_1 l_2,\,l_3 l_4}\, h_{l_3 l_4},
\label{eq:sm_dSigma}
\end{equation}
where $\lambda^{\mathrm{ch}}_{\nu,\omega}$ is the three-point vertex function defined as
\begin{align}
\lambda^{\rm ch}_{\nu\omega,\,l_1,\, l_2,\, l_3 l_4} &=
\sum_{\{l'\}} \langle c^{\phantom{*}}_{\nu,\uparrow,l'_1} c^{*}_{\nu+\omega,\uparrow,l'_2}\,\rho^{\rm ch}_{-\omega,\, l'_3 l'_4}\rangle \left[g^{-1}_{\nu}\right]_{l_1l_1'} \left[g^{-1}_{\nu+\omega}\right]_{l'_2l_2}
\end{align}
and ${\rho^{\rm ch}_{\omega,\,l_1l_2} = n^{\rm ch}_{\omega,\,l_1l_2} - \langle{}n_{\omega,\,l_1l_2}}\rangle$.
The subtraction of $\delta_{l_1,l_3}\delta_{l_2,l_4}$ removes the contribution of the bare field $h$, corresponding to the hybridization, from the self-energy, leading to the conventional form of the system’s response to external
fields~\cite{migdal1967theory}.

Reinserting the linearized self-energy response into Dyson's equation and keeping the bare action of the field gives the field-dressed propagator in a closed form
\begin{equation}
    g^h_\nu = \big[\, g_\nu^{-1} - h - \bar\lambda^{\mathrm{ch}}_{\nu,\omega=0}\, h \,\big]^{-1}
    = \big[\, g_\nu^{-1} - \lambda^{\mathrm{ch}}_{\nu,\omega=0}\, h \,\big]^{-1},
    \label{eq:sm_gh}
\end{equation}
with $g_\nu$ and $\lambda^{\mathrm{ch}}_{\nu,\omega=0}$ calculated from the dimer DMFT reference problem.
The two field-dependent terms in the first part of this relation have distinct origins: ``$-h$'' contribution comes from the shift of the hybridization function in the inverse of the bare reference Green's function, while ``$\bar\lambda^{\mathrm{ch}}_{\nu,\omega=0}\, h$'' term is the field-induced correction to the reference self-energy.
The value of the field is chosen self-consistently using the following condition:
\begin{align}
\langle{}n^{\rm ref}_{l_1l_2}[h]\rangle \overset{!}{=} \langle{}n^{\rm latt}_{l_1l_2}[h]\rangle,
\label{eq:sm_density_sc}
\end{align}
where the band indices $l_1,l_2$ in the reference and lattice densities correspond to the CDW and BOW components: the staggered diagonal components
of the densities fix the CDW field, while the off-diagonal inter-site component fixes the BOW field, which stabilizes both broken-symmetry solutions on an equal footing.

The self-consistent loop then goes as follows:\\
(i) Start from the converged dimer DMFT reference problem, calculate $g_\nu$, $\Delta_{\nu}$, and
$\lambda^{\mathrm{ch}}_{\nu,\omega=0}$, and set ${h=0}$;\\
(ii) Run the diagrammatic part of the \mbox{D-TRILEX} calculation~\cite{10.21468/SciPostPhys.13.2.036, StepanovHDR} with the current form of $g^h_\nu$, the hybridization function ${\Delta^{h}_{\nu} = \Delta_{\nu} + h}$, and the two-particle quantities of the DMFT reference problem to obtain $G_{\nu}$ and the corresponding density $n^{\mathrm{latt}}$;\\
(iii) Solve Eq.~\eqref{eq:sm_density_sc} for $h$;\\
(iv) Update $g^h_\nu$ through Eq.~\eqref{eq:sm_gh} and return to (ii).

\subsection{Application to the BOW and CDW broken-symmetry phases of the extended Hubbard model}
\label{sec:sm_application}

We now illustrate the effect of the symmetry-breaking field within the
proposed scheme, focusing on the two interaction strengths discussed in the
main text, $U=1.5$ and $U=3.0$. It is convenient to parameterize the static
field $h$ in the sublattice space of the dimer using Pauli matrices,
$h = h_x\,\sigma^x + h_z\,\sigma^z$. The identity component is fixed by the
half-filling condition, and the $\sigma^y$ component vanishes by
time-reversal symmetry.

Outside the CDW phase,
the bond-centered inversion symmetry is unbroken and forbids any staggered
on-site potential, so $h_z=0$ and the field reduces to $h=h_x\,\sigma^x$, the field conjugate to the BOW order on the intra-cluster
bond. Upon entering the CDW phase the sublattice symmetry is spontaneously
broken, and the staggered potential $h_z\,\sigma^z$, associated with the CDW order, switches on alongside the
bond field.

\begin{figure}[t!]
    \centering
    \includegraphics[width=0.95\linewidth]{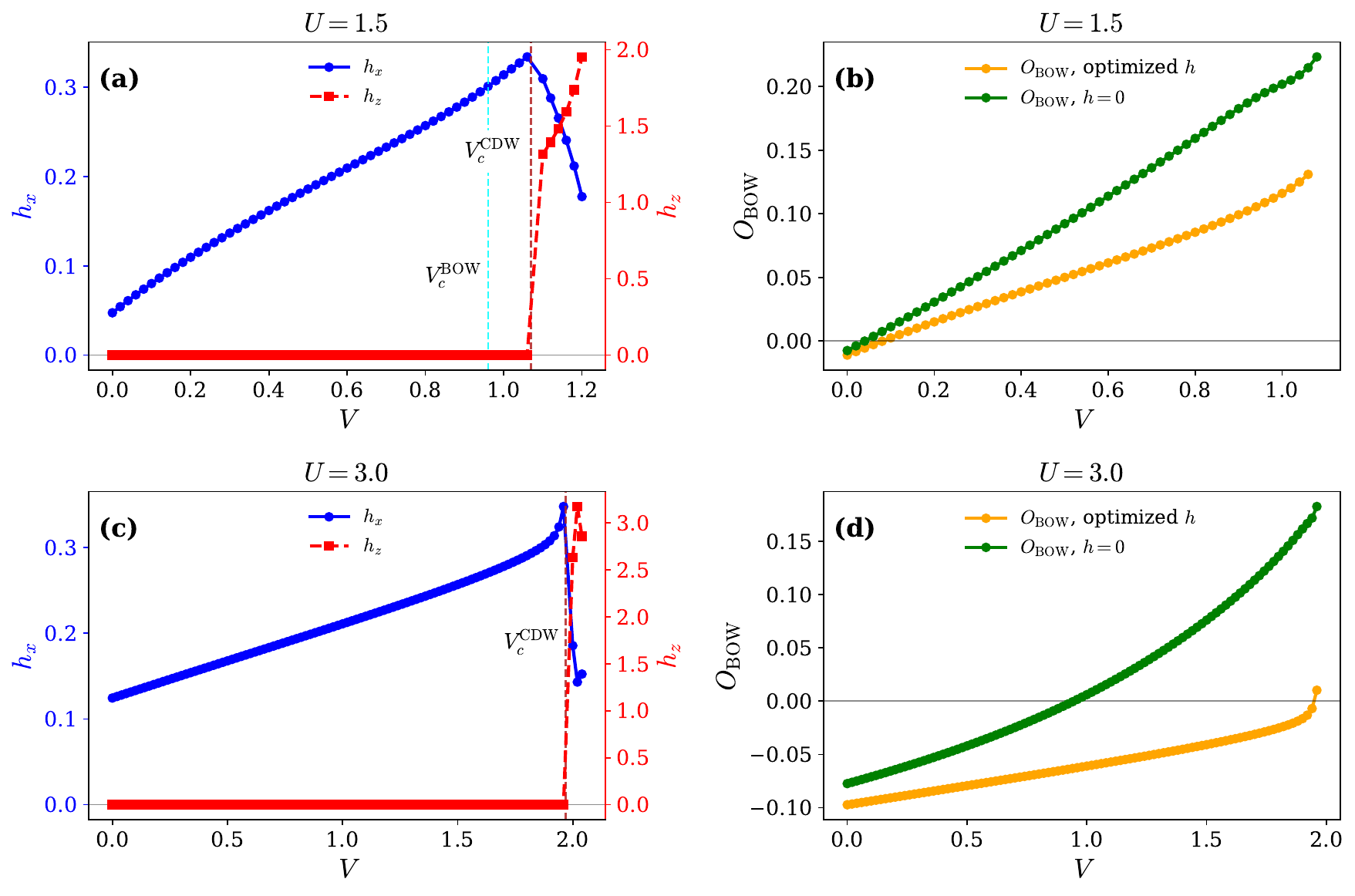}
    \caption{
        \textit{Left column:} Self-consistent local-field components versus the nonlocal interaction $V$ for (a)~$U=1.5$ and (c)~$U=3.0$.
        The bond component $h_x$ (left axis) is present throughout, while the staggered on-site component $h_z$ (right axis) activates only beyond the CDW transition.
        \textit{Right column:} BOW order parameter $O_{\rm BOW}$ versus $V$, computed with the outer self-consistency (left axis) and without it (right axis), for (b)~$U=1.5$ and (d)~$U=3.0$.
        The two are shown on separate axes because the self-consistency reduces the artifactual $O_{\rm BOW}$ background by about a factor of two. Vertical dashed lines mark $V_c^{\rm BOW}$ and $V_c^{\rm CDW}$.
    }
    \label{fig:sm_field_bow}
\end{figure}

Figure~\ref{fig:sm_field_bow}\,(a,\,c) shows how these components evolve with the
nonlocal interaction $V$. The bond field $h_x$ grows approximately linearly with $V$ in
both cases: for $U=1.5$ it rises from $h_x\simeq0.05$ at $V=0$ to
$h_x\simeq0.3$ near the CDW transition, and for $U=3.0$ from $h_x\simeq0.1$ at
$V=0$ to $h_x\simeq0.35$ at the CDW transition. This behavior is an intrinsic
feature of the scheme. The dimer DMFT reference problem corresponds to $V=0$ and therefore
accounts only for the local interaction $U$.
The nonlocal interaction $V$ is treated solely
within the \mbox{D-TRILEX} diagrammatic expansion, so as $V$ increases the dressed lattice
Green function drifts away from the reference one, and a progressively larger
$h_x$ is needed to fulfill the self-consistent condition to reconcile the two densities. 
The same logic explains the growth of $h_x$ with $U$ at fixed $V$.
This effect is visible already at $V=0$, where
$h_x$ is several times larger for $U=3.0$ than for $U=1.5$.
In one dimension, nonlocal correlations strongly affect the local quantities, and this effect increases with $U$. 
The staggered field $h_z$ behaves differently.
It is absent below the transition and turns on abruptly at $V_c^{\rm CDW}$, jumping to a finite value $h_z\approx1.3$ (for $U=1.5$) at the transition and after that rising only mildly to $h_z\approx1.6$ at $V=1.2$.

Enforcing the self-consistency brings three improvements. The first concerns
the BOW order parameter [Figure~\ref{fig:sm_field_bow}\,(b,\,d)]. 
As discussed in
the main text, the cluster geometry generates a smooth, nonzero $O_{\rm BOW}$
even in the disordered region, where no symmetry is broken. Because the field
is fixed by matching the cluster density to the more translationally
symmetric \mbox{D-TRILEX} result, it acts against this cluster-induced
asymmetry and reduces the spurious background by roughly a factor of two.

The second improvement regularizes the staggered density in the CDW phase.
Without the self-consistency condition, the symmetry-broken CDW solution is stabilized only for an unphysical value of the sublattice density difference, $\Delta n = n_A - n_B > 2$. 
Imposing the self-consistency condition~\eqref{eq:sm_density_sc} improves the reference system and yields a physically meaningful solution with the density difference constrained to the interval $\Delta n\in[0,2]$.

\begin{figure}[b!]
    \centering
    \includegraphics[width=0.6\linewidth]{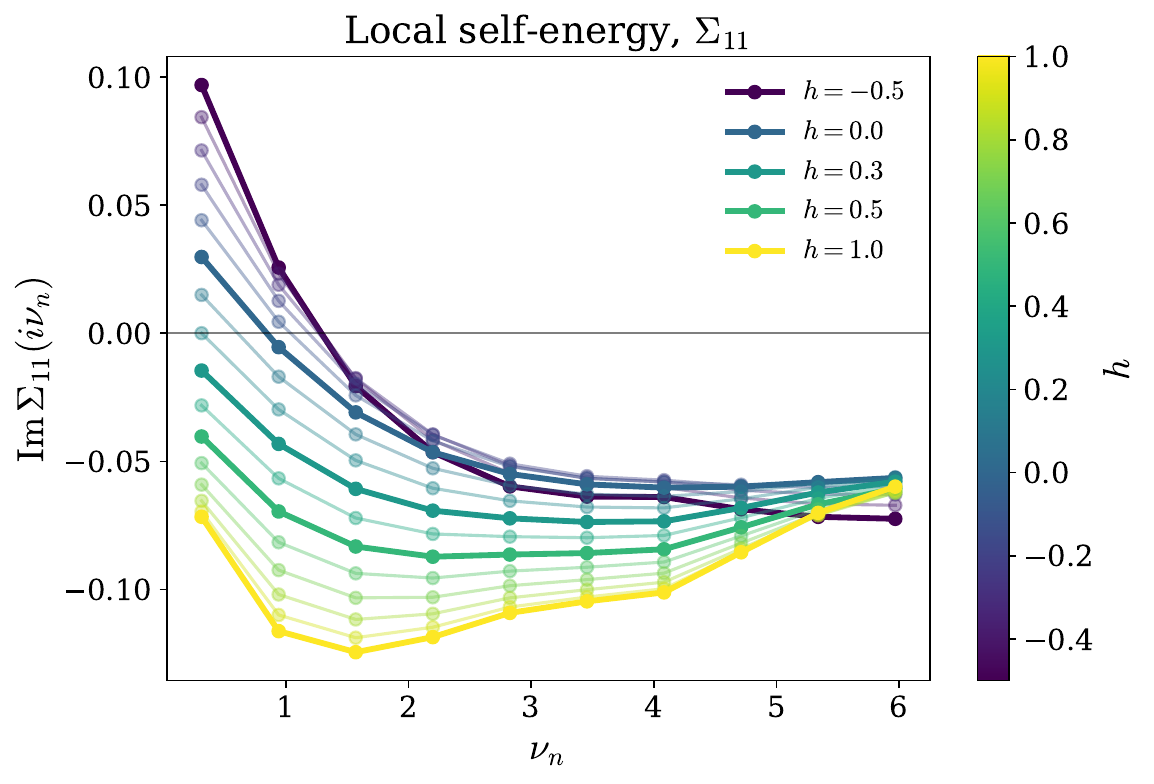}
    \caption{
        Imaginary part of the cluster self-energy
        $\mathrm{Im}\,\Sigma(\mathrm{i}\nu_n)$ versus Matsubara frequency for a
        scan of the bond field $h_x\in[-0.5,1.0]$ (color, \textit{viridis}),
        at $V=0.9$ and $U=1.5$. Curves with
        $\mathrm{Im}\,\Sigma(\mathrm{i}\nu_0)>0$ violate causality; the
        violation is cured for $h_x\in[0.2,0.3]$, where the self-energy
        extrapolates to $\mathrm{Im}\,\Sigma\to0$ as $\nu\to0^+$.
    }    \label{fig:sm_causality}
\end{figure}

The third improvement concerns causality. 
The starting point of the diagrammatic expansion is entirely determined by the hybridization function of the reference system. 
The DMFT self-consistency condition provides one possible choice for this hybridization function, but it does not guarantee that the resulting reference system remains optimal once nonlocal self-energy effects are incorporated. 
In particular, we find that for sufficiently large values of $V$, the diagrammatic expansion around the DMFT reference system obtained at $V=0$ may produce a noncausal self-energy, signaled by $\mathrm{Im}\,\Sigma(\mathrm{i}\nu_0)>0$ at the lowest Matsubara frequency.
Similar causality violations have previously been reported in other extensions of DMFT~\cite{PhysRevB.105.245115, PhysRevB.97.125141}, including the \mbox{$GW$+EDMFT}\cite{PhysRevB.95.155104, PhysRevB.104.195146, PhysRevMaterials.1.043803, PhysRevB.105.085102} and ladder D$\Gamma$A\cite{PhysRevB.91.115115, Titvinidze_2025} approaches.
Applying the approximate outer self-consistency improves the reference system by bringing it closer to the lattice solution. 
As shown in Figure~\ref{fig:sm_causality} for $U=1.5$ and $V=0.9$, increasing the symmetry-breaking field $h_x$ progressively reduces the self-energy correction and eliminates the causality violation. 
In particular, $\mathrm{Im}\,\Sigma(\mathrm{i}\nu_0)$ becomes negative for $h_x>0.2$. 
We note that the self-consistently determined value $h_x\approx0.28$ lies within this range, confirming that the proposed self-consistency procedure yields a causal solution.

\section{Topological Hamiltonian}
\label{sec:sm_topo}

\begin{figure}[t!]
    \centering
    \includegraphics[width=\linewidth]{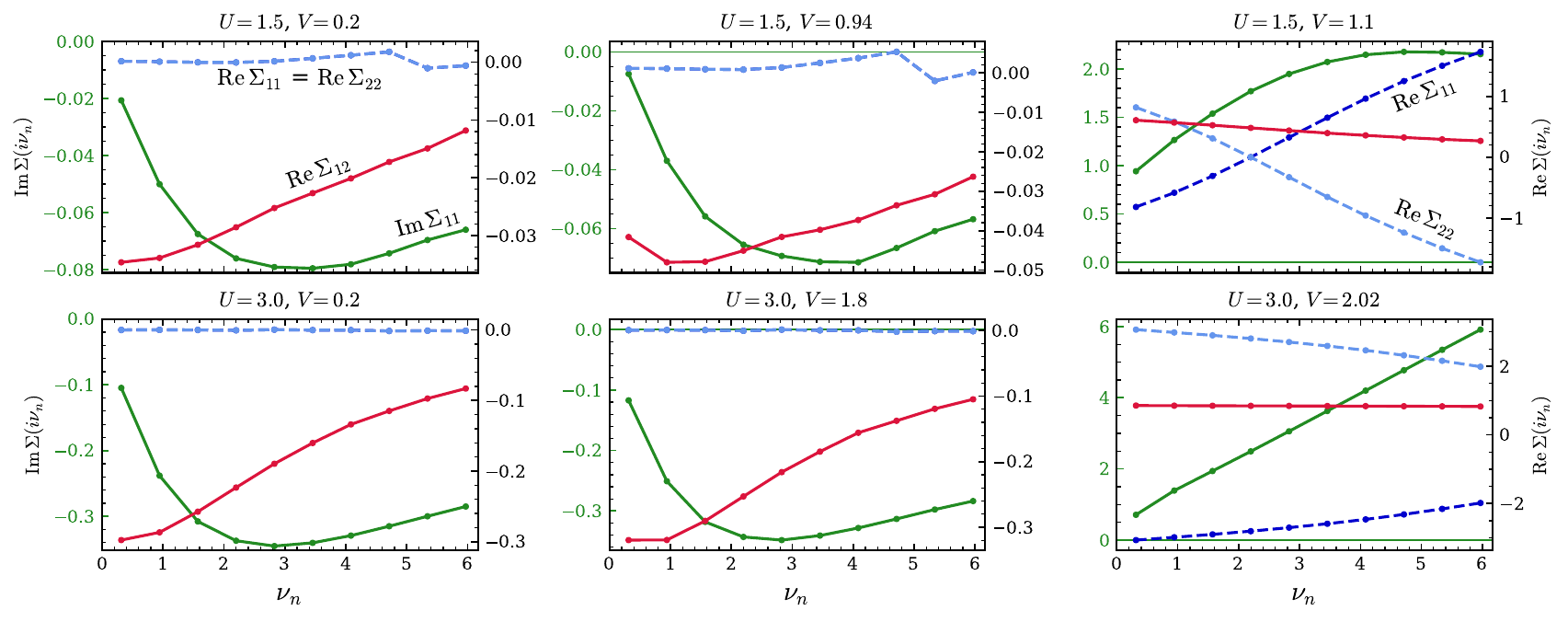}
    \caption{
        Low-frequency Matsubara self-energy $\Sigma(\mathrm{i}\nu_n)$ across
        the phases, for $U=1.5$ and $U=3.0$. The diagonal real parts
        ($\mathrm{Re}\,\Sigma_{11}$, $\mathrm{Re}\,\Sigma_{22}$) overlap in the
        normal and BOW phases but split in the CDW phase, reflecting the
        staggered potential $w_k$. The off-diagonal real part
        ($\mathrm{Re}\,\Sigma_{12}$) tracks the bond dimerization $u_k$. The
        imaginary part ($\mathrm{Im}\,\Sigma_{11}$) measures the quasiparticle
        scattering rate.
    }
    \label{fig:self_energy_panels}
\end{figure}

To determine the topological characteristics of the discovered phases, we map the
many-body problem onto an effective Hamiltonian. 
To this aim, one can perform a low-frequency expansion of the self-energy on the real-frequency axis ($\omega$):
\begin{equation}
    \Sigma(k,\omega) \simeq
    \Sigma(k,0)
    + \omega\,\partial_\omega\Sigma(k,\omega)\big|_{\omega=0}
    + \mathcal{O}(\omega^2).
\end{equation}
It allows one to write the inverse of the lattice Green's function as:
\begin{align}
G^{-1}(k,\omega) = \omega-\epsilon_{k}-\Sigma(k,\omega) \simeq
    Z^{-1}(k)\left[\omega - H_{\mathrm{eff}}(k)\right],
\end{align}
where the effective Hamiltonian $H_{\mathrm{eff}}(k) = \epsilon_{k} + \text{Re}\,\Sigma(k,0)$, following from the definition of the quasiparticle weight $Z(k) = [1-\partial_\omega\Sigma(k,\omega)|_{\omega=0}]^{-1}$.
Note that $\text{Im}\,\Sigma(k,0)=0$ holds here. 
The topological classification can be carried out directly on the effective Hamiltonian $H_{\mathrm{eff}}$, rather than on the inverse Green's function $G^{-1}(k,0)$.
While the topological invariant is a property of $G^{-1}(k,0)$, it differs only from $H_{\mathrm{eff}}$ by the factor $Z^{-1}(k)$. 
As long as $G(k,0)$ has no zeros
across the Brillouin zone, or equivalently $H_{\mathrm{eff}}$ has no poles and
$\det Z^{-1}(k)>0$, this factor is a smooth
sign-preserving deformation that does not affect the topology. 
We verify these requirements below.

To avoid the ill-conditioned analytic continuation of the data to the real frequencies, we evaluate the self-energy directly on the imaginary axis and
approximate the zero-frequency limit by the lowest fermionic Matsubara frequency, $\text{Re}\,\Sigma(k,0)\simeq\text{Re}\,\Sigma(k,\mathrm{i}\nu_0)$ with $\nu_0=\pi/\beta$.
On this basis, we evaluate the effective Hamiltonian from the dimer \mbox{D-TRILEX} self-energy to take a Bloch form in the sublattice basis:
\begin{equation}
    H_{\mathrm{eff}}(k) =
    \begin{pmatrix}
        w_k & u_k + v_k e^{-ik} \\
        u_k + v_k e^{ik} & -w_k
    \end{pmatrix},
\end{equation}
where $u_k$, $v_k$, and $w_k$ are weakly momentum-dependent parameters encoding the emergent bond and charge modulations.
Figure~\ref{fig:self_energy_panels} shows $\Sigma(\mathrm{i}\nu_n)$ at representative points across the phase diagram for $U=1.5$ and $U=3.0$, obtained by tuning $V$ through the available phases.
We find that the imaginary part of the self-energy fulfills the condition $\text{Im}\,\Sigma(k,\mathrm{i}\nu_0\to0)\to0$, which justifies the formulation of the effective Hamiltonian introduced above.

\section{Short-Range Spin Correlations and Non-Perturbative Singlet Formation}
\label{sec:sm_spin_correlations}

\begin{figure}[t!]
    \centering
    \includegraphics[width=0.7\linewidth]{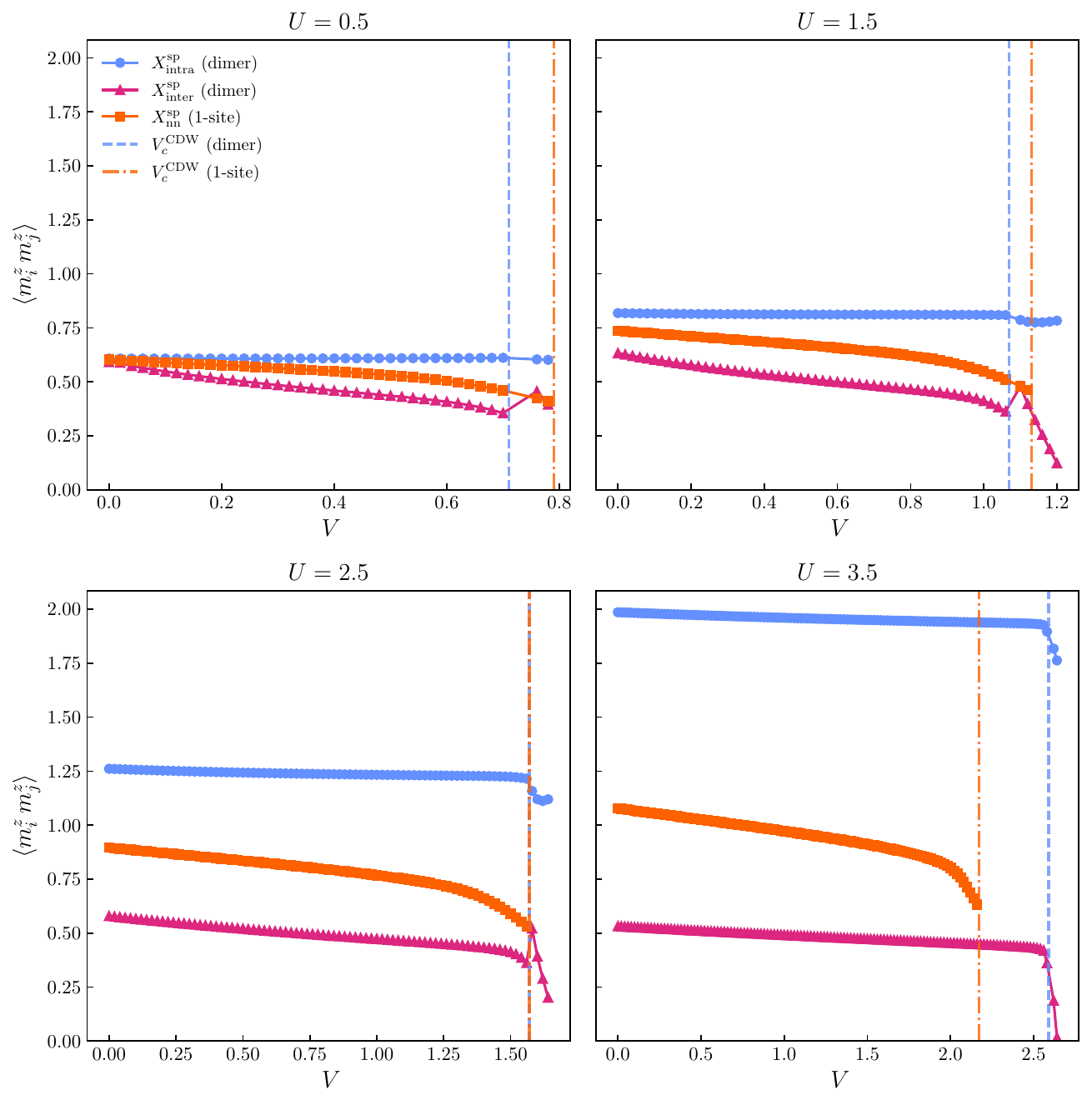}
    \caption{Nearest-neighbor spin correlations as a function of the nonlocal interaction $V$ for varying $U$. The single-site correlator $X^{\mathrm{sp}}_{\mathrm{nn}}$ (orange squares) is compared to the cluster's intra-bond $X^{\mathrm{sp}}_{\mathrm{intra}}$ (blue circles) and inter-bond $X^{\mathrm{sp}}_{\mathrm{inter}}$ (pink triangles). Vertical lines denote the respective CDW phase boundaries. At strong coupling, the cluster geometry heavily funnels correlation energy into the exactly solved intra-cluster singlet, exhibiting remarkable rigidity against $V$ until the phase transition.}
    \label{fig:spin_correlators}
\end{figure}

To understand the difference between the single-site and dimer \mbox{D-TRILEX} results for the CDW phase transition at strong coupling, we analyze the short-range spin correlations. Figure~\ref{fig:spin_correlators} presents the evolution of the nearest-neighbor static spin susceptibility ${X^{\rm sp}_{\langle{}ij\rangle} = \langle m^{z}_{i} \, m^{z}_{j} \rangle}$, where ${m^{z} = n_{\uparrow} - n_{\downarrow}}$ is the local magnetization density, as a function of $V$  for representative interaction strengths $U \in \{0.5, 1.5, 2.5, 3.5\}$. 
We compare the perturbative single-site-based result ($X^{\mathrm{sp}}_{\mathrm{nn}}$) against the non-perturbatively obtained intra-cluster ($X^{\mathrm{sp}}_{\mathrm{intra}}$) and the perturbatively treated inter-cluster ($X^{\mathrm{sp}}_{\mathrm{inter}}$) susceptibilities.
The data reveals a crossover from a perturbative regime to a regime dominated by non-perturbative singlet formation. 
At weak coupling ($U=0.5$), local moments are absent, and the magnetic fluctuations arise from electron-hole scattering. 
Consequently, all three susceptibilities are approximately the same.
As the local Coulomb repulsion increases, the system develops the local magnetic moments that, together with perturbative antiferromagnetic fluctuations, feature non-perturbative singlet correlations that are captured only by the intra-cluster susceptibility.
Consequently, the latter becomes progressively larger than the remaining two upon increasing $U$.

We note, that the single-site-based calculation explicitly enforces translational invariance and thus cannot produce bond-dependent form of the susceptibility.
For this reason, at weak to intermediate coupling $X^{\mathrm{sp}}_{\mathrm{nn}}$ lies between the intra- and inter-cluster results: $X^{\mathrm{sp}}_{\mathrm{nn}} \simeq \frac{1}{2}(X^{\mathrm{sp}}_{\mathrm{intra}} + X^{\mathrm{sp}}_{\mathrm{inter}})$.
However, in the strong-coupling case $X^{\mathrm{sp}}_{\mathrm{nn}}$ lies closer to the perturbative inter-cluster result, indicating that the non-perturbative singlet correlations are very important in this regime. 

The different behavior in these two distinct regimes is reflected in the position of the CDW phase boundary. 
Forming the CDW state requires destroying the short-range spin fluctuations, which is achieved by a nearest-neighbor Coulomb repulsion $V$.
At weak to intermediate couplings both the single-site and dimer \mbox{D-TRILEX} calculations predict a similar strength of the nearest-neighbor spin susceptibility resulting in a nearly identical value of $V_c^{\mathrm{CDW}}$.
On the contrary, the difference between the perturbative and non-perturbative susceptibilities in the strong coupling regime leads to different values of $V_c^{\mathrm{CDW}}$ predicted by the single-site and dimer calculations.
Therefore, although the cluster approach inevitably introduces an artificial bond asymmetry, it enables a non-perturbative treatment of short-range correlations, which are of crucial importance in the strong coupling regime.

\bibliography{references}